# Photostimulated spin-flip and the photodynamic therapeutic effect of fullerene solutions

Elena F. Sheka

Peoples' Friendship University of the Russia, 117198 Moscow, Russia
sheka@icp.ac.ru

Empirically estimated, fullerenes fulfill therapeutic functions acting as either antioxidant or oxidative agent thus revealing seemingly two contradictory behaviors [1]. However, this two-mode behavior is just the manifestation of the two-face appearance of fullerenes that are, on one hand, radicals due to the availability of a considerable number of effectively unpaired electrons $N_D$ [2, 3] and efficient donor-acceptor (D-A) agents [4], on the other. Actually, the consideration of chemical behavior of fullerenes discussed in [2, 3] clearly show that they must willingly interact with other radicals forming tightly bound compositions thus providing an efficient radical scavenging. In full consistence with this statement, the first exhibited therapeutic function of fullerene $C_{60}$ was its action as a radical scavenger indeed [5]. Later on this laid the foundation of the antioxidant administrating of fullerenes in medical practice [6-8]. Establishing the preservation of antioxidant properties in $C_{60}$ derivatives in general as well as its dependence on the chemical structure and, mainly, on the number of attached chemical groups with a clear preference towards monoderivatives, are in a complete accordance with expected behavior of molecular chemical susceptibility and can be quantitatively described in terms of $N_D$. It is enough to remain a clearly justified working out of this pull of effectively unpaired electrons under successive fluorination [9] and hydrogenation [10]. Therefore, the antioxidant function of fullerenes is intimately connected with electronic structure of the molecule itself.

Oppositely to individual-molecule character of the antioxidant action, the oxidative action of fullerenes occurs under photoexcitation of their solutions in both molecular and polar solvents in the presence of molecular oxygen. As accepted, the action consists in the oxidation of targets by singlet oxygen $^1O_2$ that is produced in due course of photoexcitation of fullerene solutions involving convenient triplet oxygen $^3O_2$. In the term of molecular chemical susceptibility $N_D$ within the framework of the AM1 semiempirical version of the unrestricted broken symmetry Hartree-Fock approach (UBS HF AM1) [3], the action can be explained as following. The oxygen molecule has two odd electrons that are completely engaged in the formation of the spin triplet multiplicity of the molecule in the ground state. Consequently, the total number of effectively unpaired electrons $N_D$ in this case is zero due to which convenient molecular oxygen is chemically inactive. In the singlet state, the odd electrons are exempted from the multiplicity duty and become unpaired providing $N_D$ equal $2e$ and thus exhibiting biradical character that explains $^1O_2$ high oxidative activity.

The presence of fullerene for the photostimulated $^3O_2 \rightarrow {}^1O_2$ transformation is absolutely necessary so that the treatment was called as photodynamic fullerene therapy [11, 12]. For the reason alone that the action is provided by a complex involving fullerene and solvent molecules as well as molecular oxygen, it becomes clear that it is resulted from a particular intermolecular interaction. However, until now, the mechanism of the photodynamic effect has been hidden behind a slogan 'triplet state photochemical mechanism' that implies the excitation transfer over a chain of molecules according to a widely accepted scheme [13, 14]

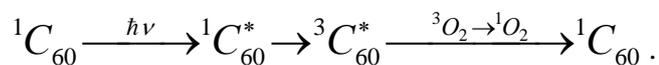

$$^1C_{60} \xrightarrow{\hbar\nu} {}^1C_{60}^* \rightarrow {}^3C_{60}^* \xrightarrow{{}^3O_2 \rightarrow {}^1O_2} {}^1C_{60} .$$

Scheme 1.

The scheme implies the energy transfer from the singlet photoexcited fullerene to the triplet one that further transfers the energy to convenient triplet oxygen thus transforming the latter into singlet oxygen. The first two stages of this 'single-fullerene-molecule' mechanism are quite evident while the third one, the most important for the final output, is quite obscure in spite of a lot of speculations available [14, 15]. Obviously, the stage efficacy depends on the strength of the intermolecular interaction between fullerene and oxygen molecules. Numerous quantum chemical calculations show that pairwise interaction in the dyad $[C_{60}+O_2]$ in both singlet and triplet states is practically zero. The UBS HF AM1 computations performed in the current work fully support the previous data and determine the coupling energy of the dyad $E_{cpl}^{f-o}$ equal zero in both cases. This puts a serious problem for the explanation of the third stage of the above scheme forcing to suggest the origination of a peculiar intermolecular interaction between $C_{60}$ and $O_2$ molecules in the excited state once absent in the ground state.

However, the intermolecular interaction in the photodynamic (PD) solutions is not limited by the fullerene-oxygen (*f-o*) interaction only. There are two other interactions, namely: fullerene-fullerene (*f-f*) and fullerene-solvent (*f-s*), among which the former is quite significant thus revealing itself in a considerable amplification of the spectral properties of fullerene solutions [16]. The *f-s* interaction in the case of aqueous and benzene solutions can be ignored. Besides a significant strength, the *f-f* interaction possesses some peculiar features caused by the exclusive D-A ability of fullerenes. As we know from a detailed consideration of the fullerene dimerization [17], a significant contribution of the D-A component into the total intermolecular interaction results in a two-well shape of the potential energy term of a pair of fullerene molecules in the ground state, which is schematically shown in Fig.1. According to the scheme, the pairwise interaction between the molecules in convenient solutions always leads to the formation of bi-molecular or more complex homoclusters of fullerenes in the vicinity of the $R^{00}$ minimum on the potential energy curve. The dimerization (as well as oligomerization) is a barrier reaction and does not occur spontaneously. Particular measures should be undertaken to come over the barrier and provide the molecule chemical coupling whilst the dimerization is energetically profitable. And photoexcitation is one of the most efficient tools. Therefore the PD solutions under ambient conditions should involve conglomerates of clusterized $C_{60}$ molecules as shown schematically in Fig.2, which is experimentally proven in many cases (see for example [16, 18, 19]).

UBS HF AM1 calculations determine the coupling energy of the pairwise *f-f* interaction for $C_{60}$ as $E_{cpl}^{f-f}$=-0.52 *kcal/mol*. If remember that $E_{cpl}^{f-o}$=0 in both singlet and triplet state, it becomes clear that oxygen molecules do not interact with either individual fullerene molecule or the molecule clusters so that the total energy of any dyad $[(C_{60})_n - O_2]$ ( n=1, 2, 3….) is just a sum of those related to the dyad components and is always by 9.93 kcal/mol less in the triplet state due to the difference in the energy of the triplet and singlet oxygen (the UBS HF AM1 energy of $^3O_2$ and $^1O_2$ molecules constitutes -27.75 and -17.82 kcal/mol, respectively). Therefore the ground state of the dyads is triplet.

Computations have shown [4, 16, 17] that each pair of fullerene molecules as well as any fullerene cluster of more complex structure formed at the $R^{00}$ minimum are charge transfer complexes. Their absorption bands related to $B_2$ phototransitions in Fig.1 are located in the UV-visible region. The photoexcitation of either pair or cluster of fullerene molecules within this region produces a pair of molecular ions that quickly relax into the ground state of neutral molecule after the light is switched off. The calculations have revealed that, oppositely to neutral $C_{60}$, both molecular ions $C_{60}^-$ and $C_{60}^+$ actively interact with oxygen molecule producing coupling energy $E_{cpl}^{-f-o}$ and $E_{cpl}^{+f-o}$ of -10.03 and -10.05 *kcal/mol*, respectively, referring to $^3O_2$ molecule and -0.097 and -0.115 *kcal/mol* in regards to $^1O_2$. Therefore, oxygen molecule is quite strongly held in the vicinity of both molecular ions forming $[C_{60}+O_2]^-$ and

$[C_{60}+O_2]^+$ complexes as schematically shown in Fig.3. UBS HF AM1 calculations for the corresponding dyads show that the complexes are of $^2[C_{60}^-+O_2]$ and $^2[C_{60}^++O_2]$ compositions of the doublet spin multiplicity. Both fullerene ions take the responsibility over the complex spin multiplicity, so that two odd electrons of the oxygen molecule are not more to maintain the molecule triplet multiplicity and become unpaired thus adding two effectively unpaired electrons to the $N_D$ pool of unpaired electrons of the whole complex. The distribution of unpaired electrons of both complexes over their atoms, which displays the distribution of the atomic chemical susceptibility of the complexes, is shown in Fig.4. A dominant contribution of electrons located on oxygen atoms 61 and 62 is clearly seen thus revealing the most active sites of the complexes. It should be noted that these distributions are intimate characteristics of both complexes so that not oxygen itself but $^2[C_{60}^-+O_2]$ and $^2[C_{60}^++O_2]$ complexes as a whole provide the oxidative effect. The effect is lasted until the complexes exist and is practically immediately terminated when the complexes disappear when the light is switched off.

The obtained results make it possible to suggest the following mechanism that lays the foundation of the photodynamic effect of fullerene solutions

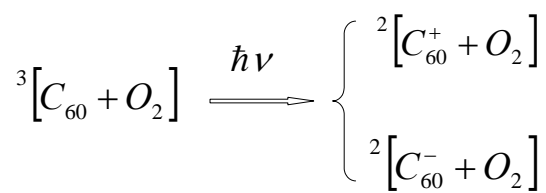

Scheme 2

The corresponding atomic compositions are shown in Fig.5. As shown in the figure, changing spin multiplicity from the triplet to doublet one under photoexcitation due to passing from neutral molecule complex to those based on fullerene molecular results in a spin flip in the system of two odd electrons of the oxygen molecule. This approach allows attributing phodynamical effect of fullerene solutions to a new type of chemical reactions in the modern spin chemistry.

Since fullerene derivatives preserve D-A properties of the pristine fullerene, Scheme 2 is fully attributable to the latter as well. So that not only $C_{60}$ or $C_{70}$ themselves but their derivatives can be used in PD solutions. Obviously, parameters of the photodynamic therapy should therewith be different depending on the fullerene derivative structure as is actually observed experimentally [15]. Changing solute molecules, it is possible to influence the efficacy of their clusterization, which, in its turn, may either enhance or press the therapeutic effect. The situation appears to be similar to that occurred in nanophotonics of fullerene solution [17].

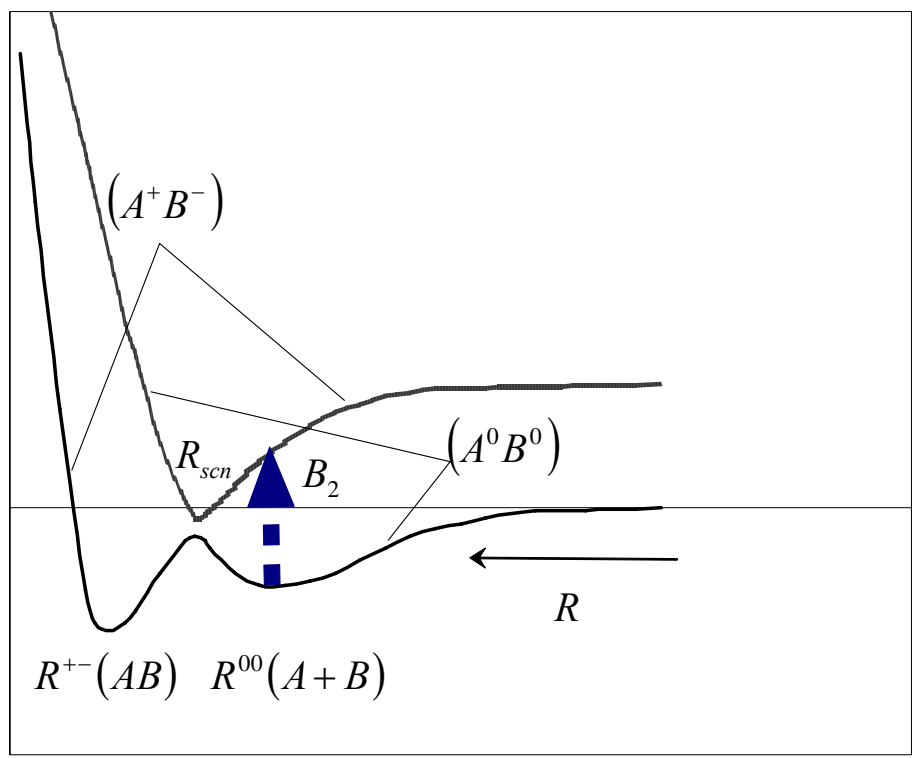

**Figure 1**. Scheme of intermolecular-interaction terms related to the interaction between two fullerene molecules [17]. $(A^0 B^0)$ and $(A^+ B^-)$ match branches of the terms related to the interaction between neutral molecules and their ions, respectively. $R^{+-}$ and $R^{00}$ mark minimum positions attributed to the formation of tightly bound dimer (AB) and weakly bound charge transfer complex (A+B), respectively. $R_{scn}$ indicates the point of avoidable intersection of terms $(A^0 B^0)$ and $(A^+ B^-)$.

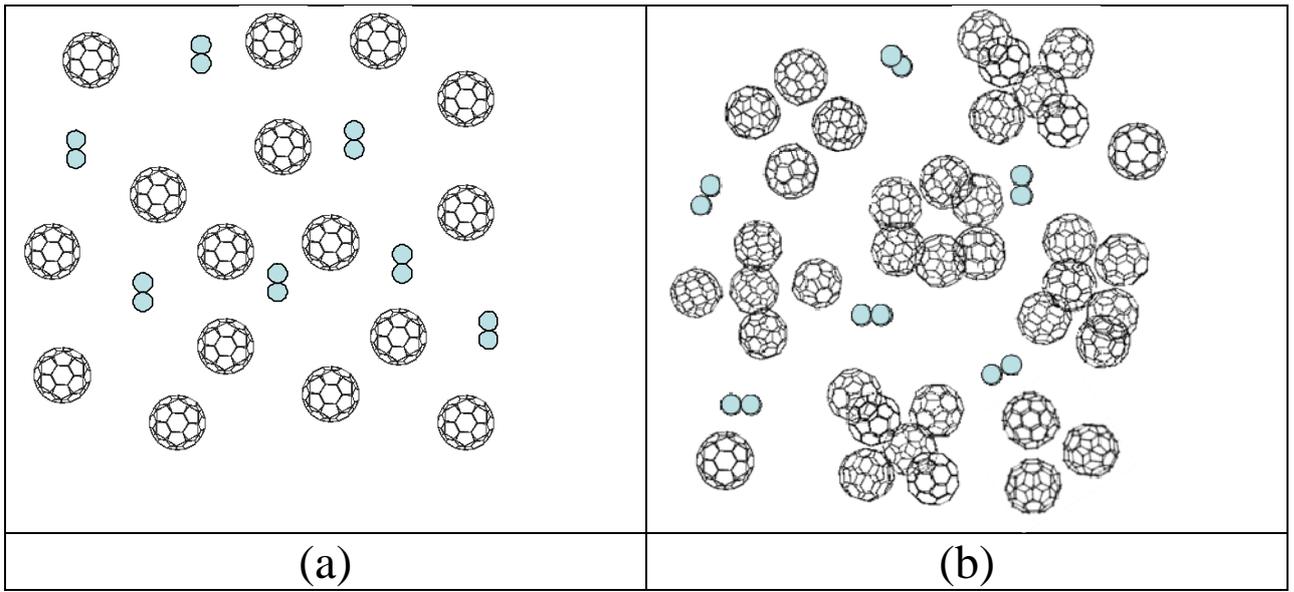

**Figure 2.** A schematic presentation of an ideal (*a*) and real (*b*) fullerene solution

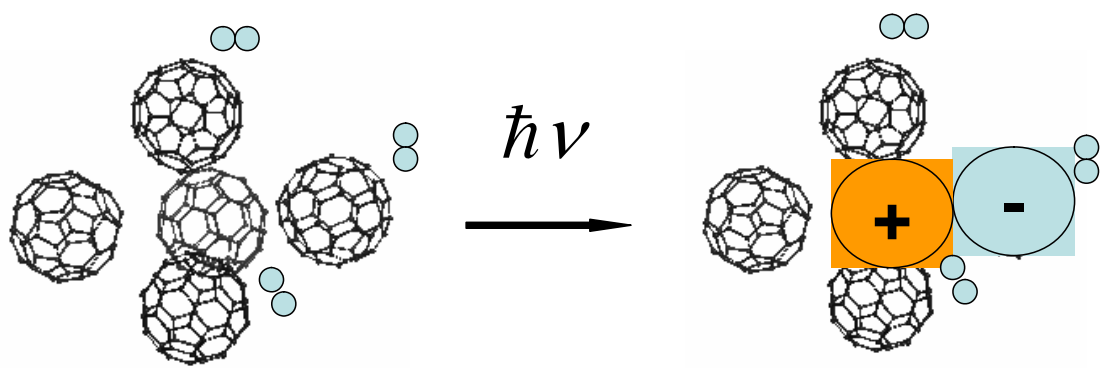

**Figure 3.** The formation of an ionic pair of $C_{60}$ under photoexcitation.

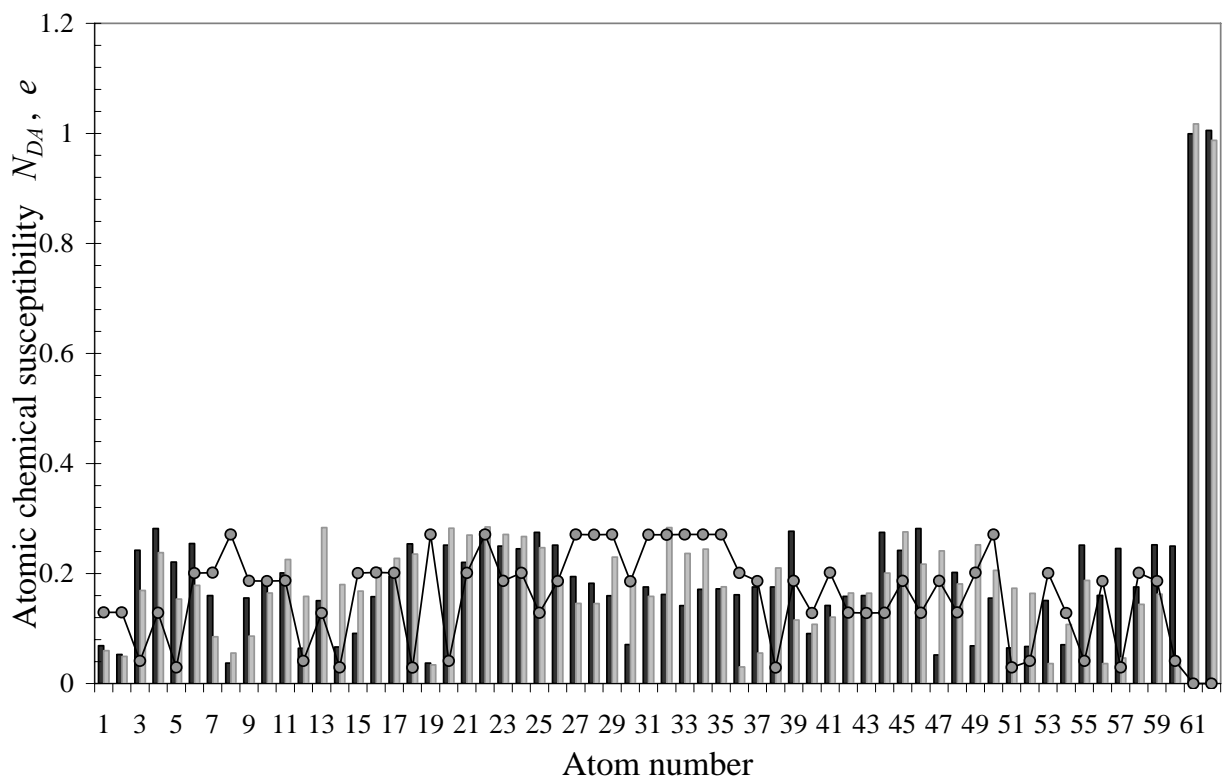

**Figure 4.** Distribution of atomic chemical susceptibility $N_{DA}$ [3] over atoms of $^2[C_{60}^- + O_2]$ (black bars) and $^2[C_{60}^+ + O_2]$ (light gray bars) complexes. Curve with black dots plots the distribution over atoms of $^3[C_{60} + O_2]$ complex.

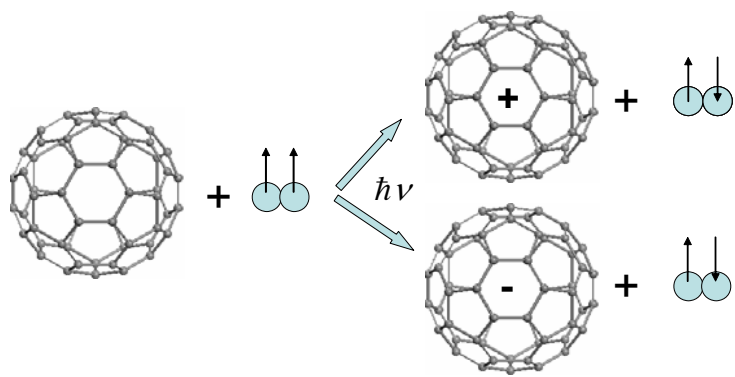

**Figure 5**. A schematic presentation of the spin-flip in oxygen molecule under photoexcitation